\documentclass[letterpaper]{article} 
\usepackage{natbib}
\usepackage{2026}  
\usepackage{times}  
\usepackage{helvet}  
\usepackage{courier}  
\usepackage[hyphens]{url}  
\usepackage{graphicx} 
\usepackage{natbib}  
\usepackage{caption} 
\usepackage{enumitem}
\usepackage{amsmath} 
\usepackage{booktabs}
 \usepackage{multirow}
\usepackage{cleveref}
\crefname{section}{Sec.}{Secs.}
\Crefname{section}{Sec.}{Secs.}
\usepackage{amsfonts}

\usepackage{algorithm}
\usepackage{algorithmic}

\usepackage{newfloat}
\usepackage{listings}
\DeclareCaptionStyle{ruled}{labelfont=normalfont,labelsep=colon,strut=off} 
\floatstyle{ruled}
\newfloat{listing}{tb}{lst}{}
\floatname{listing}{Listing}
\nocopyright 
\title{RLGS: Reinforcement Learning‑Based Adaptive Hyperparameter Tuning for Gaussian Splatting}
\author{
    Zhan Li, Huangying Zhan, Changyang Li, Qingan Yan, Yi Xu}
\affiliations{
    Goertek Alpha Labs\\
   \texttt{\{first.last\}@goertekusa.com}\\

}

\begin{document}

\makeatletter
\g@addto@macro\@maketitle{
  \includegraphics[width=1.0\textwidth]{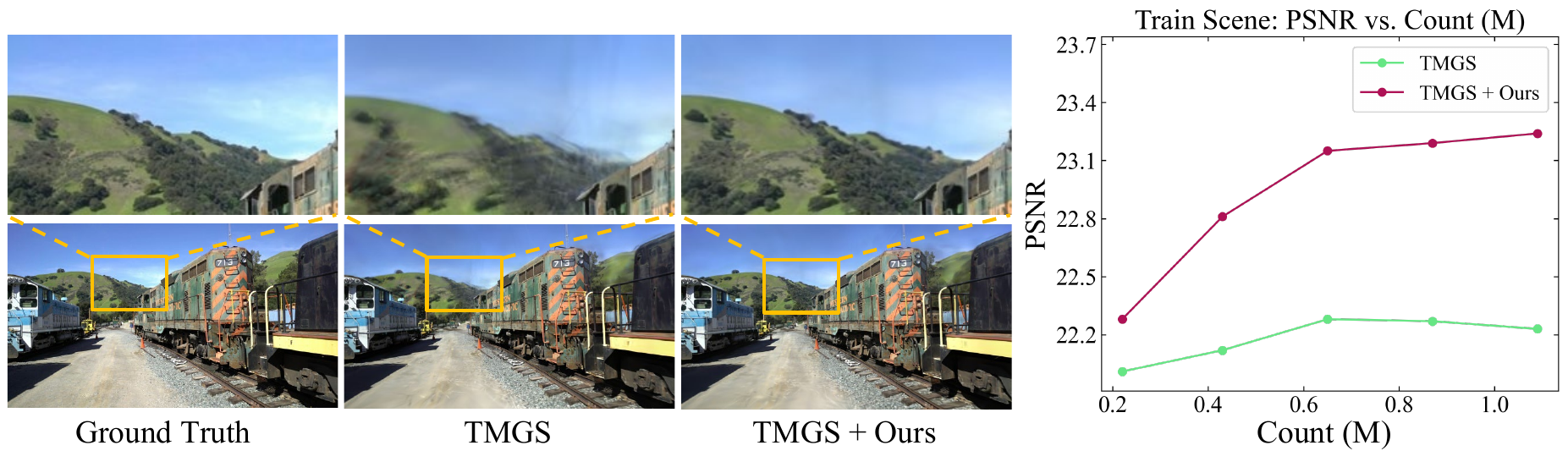}
  \vspace{-5mm}
  \captionof{figure}{Our reinforcement learning hyperparameter tuning boosts the state‑of‑the‑art Taming 3DGS (TMGS) across varying number of Gaussians. In the Train scene where adding Gaussians fails after 0.7 million, our method still scales up quality.}
  \vspace{-2mm}
  \label{fig:teaser}\vspace{15pt}
}
\makeatother
\maketitle

\begin{abstract}
Hyperparameter tuning in 3D Gaussian Splatting (3DGS) is a labor-intensive and expert-driven process, often resulting in inconsistent reconstructions and suboptimal results. 
We propose RLGS, a plug-and-play reinforcement learning framework for adaptive  hyperparameter tuning in 3DGS through lightweight policy modules, dynamically adjusting critical hyperparameters such as learning rates and densification thresholds.
The framework is model-agnostic and seamlessly integrates into existing 3DGS pipelines without architectural modifications.
We demonstrate its generalization ability across multiple state-of-the-art 3DGS variants, including Taming-3DGS and 3DGS-MCMC, and validate its robustness across diverse datasets.
RLGS consistently enhances rendering quality.
For example, it improves Taming-3DGS by \textbf{0.7dB} PSNR on the Tanks and Temple (TNT) dataset, under a fixed Gaussian budget, and continues to yield gains even when baseline performance saturates.
Our results suggest that RLGS provides an effective and general solution for automating hyperparameter tuning in 3DGS training, bridging a gap in applying reinforcement learning to 3DGS.

\end{abstract}

\section{Introduction}

Photorealistic novel view synthesis (NVS) is a foundational problem in computer vision and graphics, with applications spanning immersive virtual environments and synthetic data generation for autonomous systems.
Recent advances in neural representations have significantly improved both rendering fidelity and generalization.
Among NVS methods, neural radiance fields (NeRFs)~\cite{mildenhall2020nerf} introduced a paradigm shift by modeling scenes as continuous neural functions.
However, this continuous volumetric representation incurs a substantial computational cost during both training and inference.
To address these limitations, 3D Gaussian Splatting~\cite{kerbl3Dgaussians} has emerged as a highly efficient alternative, enabling real-time rendering through a point-based scene representation and GPU-friendly rasterization.

Despite its promising performance, 3DGS remains highly sensitive to hyperparameter settings, such as learning rates and densification thresholds.
These parameters govern reconstruction quality, but selecting appropriate values remains a tedious, manual, and scene-specific process.
In practice, even slight mis-configurations can lead to suboptimal rendering, overfitting, or excessive Gaussian growth, significantly limiting the accessibility and scalability of 3DGS in real-world pipelines.

While hyperparameter optimization (HPO) is a well-established topic in machine learning~\cite{bergstra2012random, snoek2012practical, li2018hyperband, jaderberg2017population}, most existing methods, such as grid search, Bayesian optimization, and population-based training, are not suited to the demands of 3DGS.
These approaches typically assume access to low-cost evaluations or differentiable objectives, and often require dozens of full training runs.
This renders them impractical for computationally intensive tasks like 3DGS, where each trial can take tens of minutes and hyperparameters interact nontrivially with dynamic scene evolution.

We propose a novel solution: casting hyperparameter control in 3DGS as an online decision-making problem and addressing it with reinforcement learning.
Specifically, we introduce a policy-gradient framework that learns to dynamically adjust hyperparameters, such as learning rates and densification thresholds, during training of 3DGS models.
Our method treats 3DGS optimization as a Markov Decision Process (MDP), where a lightweight policy observes training progress and photometric error, and outputs scaling factors for key hyperparameters.

Our approach enables fine-grained, scene-aware adaptation of hyperparameters, eliminates manual tuning, and integrates seamlessly into existing 3DGS pipelines. 
Notably, in scenes where state-of-the-art methods plateau despite increasing the number of Gaussians, our method continues to improve rendering quality, as illustrated in Figure~\ref{fig:teaser}.
We summarize our main contributions as follows:

First, we are the first to formulate hyperparameter tuning for 3D Gaussian Splatting as a reinforcement learning problem. Our framework learns adaptive policies that dynamically adjust hyperparameters during training to enhance rendering quality. 

Second, we introduce a lightweight, modular architecture where separate policy agents independently control learning rate schedules and densification behavior. This design enables targeted, interpretable adaptation over different stages of 3DGS training.

Finally, we validate RLGS across two state-of-the-art 3DGS variants: Taming-3DGS and 3DGS-MCMC, and demonstrate consistent improvements on multiple benchmarking datasets.

\section{Related Work}
We review prior work in five key areas relevant to our method: novel view synthesis and neural representations, 3D Gaussian Splatting, hyperparameter optimization, reinforcement learning for training control, and the integration of reinforcement learning within 3DGS systems.

\paragraph{Novel View Synthesis and Neural Representations.}
Novel view synthesis (NVS) aims to generate unseen views of a scene from a set of input images.
Early methods~\cite{Debevec96,gortler1996lumigraph,levoy1996light,Heigl1999,Buehler2001,zheng2009parallax,kopf2014first,ortiz2015bayesian,Hedman2016,Penner2017,hedman2018deep,flynn2019deepview,wiles2020synsin} relied on image-based rendering techniques with proxy geometry or depth maps to approximate novel views.
These approaches benefit from strong geometric priors but struggle to generalize to complex scenes with significant occlusions or view-dependent effects.
Recent neural approaches replace explicit geometry with learnable representations, such as volumetric grids~\cite{sitzmann2019deepvoxels} or implicit neural fields~\cite{srn_sitzmann_2019}, offering better generalization at the cost of higher computational demands.
Neural Radiance Fields (NeRF)~\cite{mildenhall2020nerf,barron2023zipnerf} marked a major advance by modeling the scene as a continuous 5D radiance field, producing photorealistic renderings but requiring long training and inference times.

\paragraph{3D Gaussian Splatting.}
3D Gaussian Splatting (3DGS)~\cite{kerbl3Dgaussians} offers a fast and scalable alternative to NeRF by representing scenes as a set of anisotropic 3D Gaussians, rendered using efficient rasterization instead of ray marching.
Subsequent research has extended 3DGS in several directions.
For improved geometry, methods such as~\cite{Huang2DGS2024,yu2024gaussian} refine surface fitting and better capture fine details.
For rendering quality, Mip-Splatting~\cite{Yu2024MipSplatting} fuses multi-resolution Gaussians to reduce aliasing.
To scale to large scenes, hierarchical~\cite{kerbl2024hierarchical} and block-partitioned~\cite{lin2024vastgaussian} structures are introduced.
Parameter compression is achieved via quantization and sparsification~\cite{papantonakis2024reducing,fan2024lightgaussian,mallick2024taming}. Recent works also focus on rendering improvements~\cite{mai2024ever,hou2024sort,celarek2025does,kheradmand2025stochasticsplats} and optimization algorithms~\cite{hollein20243dgs,lan20253dgs2nearsecondorderconverging,pehlivan2025second}. Notably, Taming-3DGS~\cite{mallick2024taming} introduces error-aware pruning and learned importance metrics to remove low-impact Gaussians and control memory overhead without sacrificing visual fidelity. 3DGS-MCMC~\cite{kheradmand20243d} integrates a Markov Chain Monte Carlo (MCMC) procedure into the 3DGS optimization loop, treating Gaussian parameters as random variables and iteratively sampling proposals to refine the scene representation. However, hyperparameters such as learning rates and pruning thresholds in these variants are still manually tuned by experts, and none of these works offer an automated tuning solution.

\paragraph{Hyperparameter Optimization.}
Hyperparameter optimization (HPO) is crucial for achieving optimal performance in deep learning models.
Traditional methods such as grid search and random search~\cite{bergstra2012random} are simple but computationally inefficient.
More advanced techniques include Bayesian optimization~\cite{snoek2012practical}, Hyperband~\cite{li2018hyperband}, and population-based training (PBT)~\cite{jaderberg2017population}, which balance exploration and resource efficiency.
These approaches have been applied to tune neural architectures, training schedules, and loss weights across various domains.
However, they are typically designed for offline, trial-based search and are not well-suited for pipelines like 3DGS.

\paragraph{Reinforcement Learning for Optimization and Control.}
Reinforcement learning (RL) has emerged as a powerful tool for automating control and optimization in machine learning pipelines. It has been applied to neural architecture search~\cite{zoph2017nas,pham2018efficient}, adaptive learning rate scheduling~\cite{daniel2016learning,wu2018variance}, and curriculum learning~\cite{graves2017automated}. Policy-gradient methods, such as REINFORCE~\cite{williams1992simple}, are especially effective in settings involving sparse or delayed rewards and are widely used in non-differentiable or black-box optimization problems. Recent work has also explored meta-RL approaches for learning optimizers~\cite{andrychowicz2016learning, xu2018meta}. These methods have demonstrated the potential of RL to discover dynamic training strategies that outperform static heuristics.

\paragraph{Reinforcement Learning for 3DGS}
Several recent works~\cite{wang2024reinforcement,wang2024query,wu2024rl} integrate 3DGS into RL environments for downstream tasks such as robotic control or semantic exploration.
These efforts treat 3DGS as a world model or rendering engine to support learning agents.
In contrast, we use reinforcement learning not as a consumer of 3DGS outputs, but as a controller that directly improves the training process of 3DGS.

To the best of our knowledge, this is the first work to apply reinforcement learning for online hyperparameter optimization in 3D Gaussian Splatting.
We address a key gap in the literature by introducing a learning-based controller that adaptively tunes hyperparameters during training to improve rendering quality and efficiency.

\begin{figure*}[ht]
  \centering
  \includegraphics[width=1.0\linewidth]{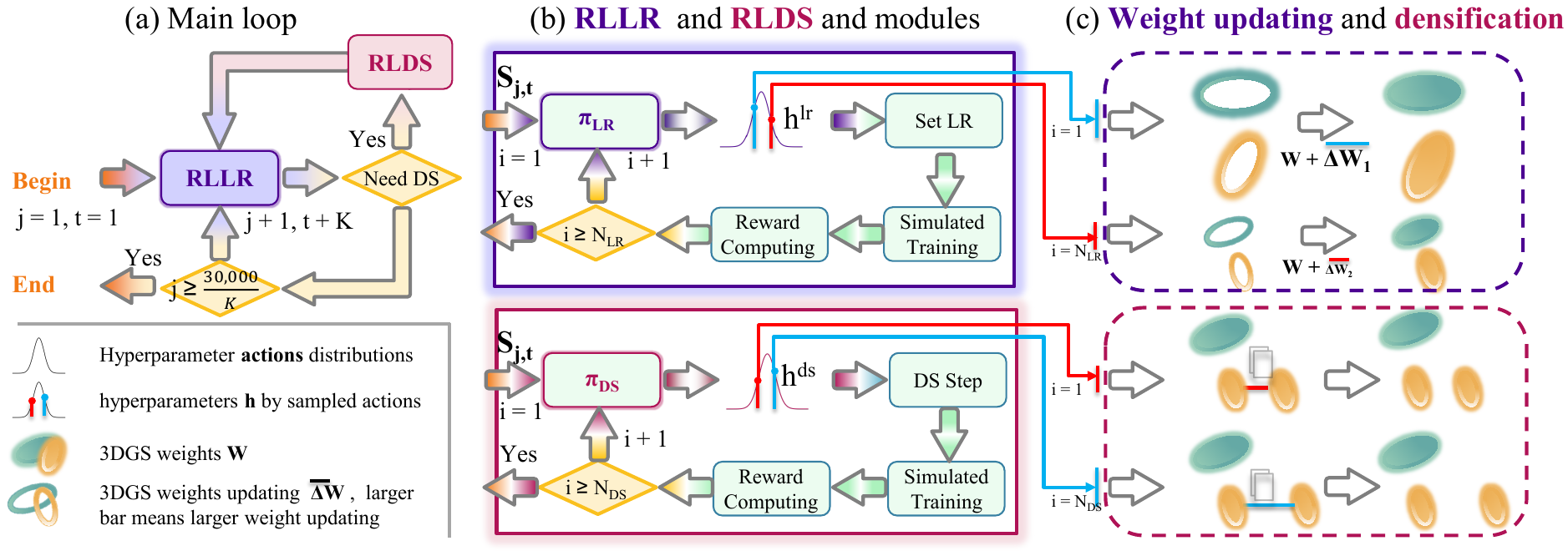}
  \caption{\textbf{Overview of the proposed framework}. 
    (a) Our method follows the vanilla 3DGS training loop with two plug-and-play modules, $\mathrm{RLLR}$ and $\mathrm{RLDS}$, for adjusting the learning rate and densification hyperparameters, respectively. (b) This sub-figure presents our $\mathrm{RLLR}$ and $\mathrm{RLDS}$ modules. At policy phase $j$ with global iteration $t$, the two policies take as input the state $s_{j,t}$, which encodes the number of completed training iterations and the previous phase’s loss (see Equation~\ref{eq:input}). Each policy then samples an action, $a_{j,t}^{\mathrm{lr}}$ and $a_{j,t}^{\mathrm{ds}}$, respectively, to generate the hyperparameters $h^{\mathrm{lr}}$ and $h^{\mathrm{ds}}$. Both policies are optimized via policy gradient to maximize reconstruction improvement after simulated training. (c) This sub-figure illustrates how hyperparameters $h^{\mathrm{lr}}$ and $h^{\mathrm{ds}}$ control updates to the model weights $\mathrm{W}$ and affect densification, respectively.}
  \label{fig:framework}
\end{figure*}

\section{Methodology}
We treat hyperparameter tuning in 3DGS as a reinforcement learning problem, where lightweight policy modules dynamically adjust key hyperparameters during training. To this end, we propose two lightweight policy modules—RLLR for adaptive learning‐rate scaling (policy~$\pi_{\mathrm{LR}}$) and RLDS for densification adjustment (policy~$\pi_{\mathrm{DS}}$). 
As shown in Figure~\ref{fig:framework}, these two modules are plugged into the 3DGS training pipeline and can be integrated into other 3DGS variants.

\subsection{Problem Formulation}
\label{sec:problemformulation}
We formulate hyperparameter tuning as a reinforcement-learning task, using $\mathrm{RLLR}$ to adapt the learning rate and ${\mathrm{RLDS}}$ to adjust densification hyperparameters when needed. To mitigate training variance in 3DGS that hurts policy optimization, we fold the original 
30{,}000-step 
3DGS training schedule into 
$J = \frac{30{,}000}{K}$
policy-phases, each spanning $K$ training steps of 3DGS. We also maintain a global iteration $t$ to represent the training progress.

During policy-phase $j$, ${\mathrm{RLLR}}$ executes an inner policy update loop with maximum length of $N_{lr}$ steps. The policy observes state $s_{j,t}$ as follows:
\begin{equation}\label{eq:input}
  s_{j,t} = \bigl(\hat{\ell}_{j-1},\,\hat{\tau}_{t}\bigr)
\end{equation}
where $\hat{\ell}_{j-1}$ denotes the previous policy-phase's 3DGS training loss and $\hat{\tau}_{t}$ encodes the current training iteration $t$. The policy outputs scaling factors $a^{\mathrm{lr}}_{j,t}$ and uses them to adjust the original learning rate $h^{\mathrm{lr}}_{\mathrm{orig}}$ as follows:
\begin{equation}
  a_{j,t}^{\mathrm{lr}} \sim \pi_{\mathrm{LR}}\bigl(\,\cdot\,\bigm|s_{j,t}\bigr),
   \quad
  h^{\mathrm{lr}} = h^{\mathrm{lr}}_{\mathrm{orig}} \odot a^{\mathrm{lr}}_{j,t}
\end{equation}

If densification is required, the ${\mathrm{RLDS}}$ module is also invoked using the same state input, and produces scaling factors $a^{\mathrm{ds}}_{j,t}$ for densification:
\begin{equation}
   a_{j,t}^{\mathrm{ds}} \sim \pi_{\mathrm{DS}}\bigl(\,\cdot\,\bigm|s_{j,t}\bigr),
   \quad
    h^{\mathrm{ds}} = h^{\mathrm{ds}}_{\mathrm{orig}} \odot a^{\mathrm{ds}}_{j,t}.
\end{equation}

Each inner iteration of ${\mathrm{RLLR}}$ consists of sampling learning rates, applying them for $K$ simulated training steps, computing the reconstruction improvement reward $R_{j,t}^{\mathrm{LR}}$, and updating the policy. Similarly, each inner iteration of ${\mathrm{RLDS}}$ samples densification parameters, performs a densification step, trains for $K$ simulated steps, computes $R_{j,t}^{\mathrm{DS}}$, and updates the densification policy.

The reward is defined as the improvement over the default hyperparameters $h_{\mathrm{orig}}$:
\begin{equation}
\label{eq:reward}
  R_{j,t} = \mathcal{M}(h) - \mathcal{M}(h_{\mathrm{orig}})
\end{equation}where $\mathcal{M}(\cdot)$ denotes a rendering error metric evaluated on reward views.

After $N_{\mathrm{LR}}$ inner iterations for RLLR and $N_{\mathrm{DS}}$ for RLDS, we select the best-performing configuration $(h^{\mathrm{lr}}, h^{\mathrm{ds}})$ and apply it to the actual update of the current policy-phase~$j$. Repeating this process over 
$30{,}000/K$ 
policy-phases yields an efficient RL framework that dynamically adapts the learning rate and densification parameters to maximize reconstruction quality while preserving the original 3DGS pipeline.

\subsubsection{Network Architecture}
We use same network architecture for both policies. Assume $d$ is the number of hyperparameters that need to be tuned in $\mathrm{RLLR}$ or $\mathrm{RLDS}$. Each policy module contains a neural network to predict the distribution of hyperparameters. Specifically, the network consists of a GRU cell encoder~\cite{cho2014properties} and a linear head for each hyperparameter. The network takes input state $s_{j,t}$ as in~\ref{eq:input} and produces residual outputs $\Delta\mu \in \mathbb{R}^{d}$ and $\Delta\log\sigma \in \mathbb{R}^{d}$. In addition, the module maintains two learned base parameters, $\mu_{\mathrm{base}}\in \mathbb{R}^{d}$ and $\log\sigma_{\mathrm{base}} \in \mathbb{R}^{d}$. The final mean and log standard deviation are obtained by adding the network-predicted residuals to these learnable bases:
\begin{equation}
\begin{aligned}
\mu &= \mu_{\mathrm{base}} + \Delta\mu,\\
\log\sigma &= \log\sigma_{\mathrm{base}} + \Delta\log\sigma.
\end{aligned}
\end{equation}

\subsubsection{Action Sampling and Exploration}
An action $a \in \mathbb{R}^{d}$ is sampled from the resulting Gaussian distribution:
\begin{equation}
a \sim \mathcal{N}\bigl(\mu,\;\sigma^2\bigr),
\end{equation}
where $\sigma = \exp(\log\sigma)$. To encourage exploration, we add an entropy bonus to the policy objective \cite{mnih2016asynchronous,schulman2017proximal}. We train the policy by minimizing the negative reward-weighted log-probability of actions and an entropy term.

\subsection{Reward Design}

\subsubsection{Simulated Training}
The effect of a sampled hyperparameter action $a$ appears only after several training iterations. After sampling a hyperparameter action $a$, we simulate the future $K$ training iterations to compute the reward as Equation~\ref{eq:reward}.

\subsubsection{Reward View Sampling}
\begin{figure}[t]
  \centering
  \includegraphics[width=1.0\linewidth]{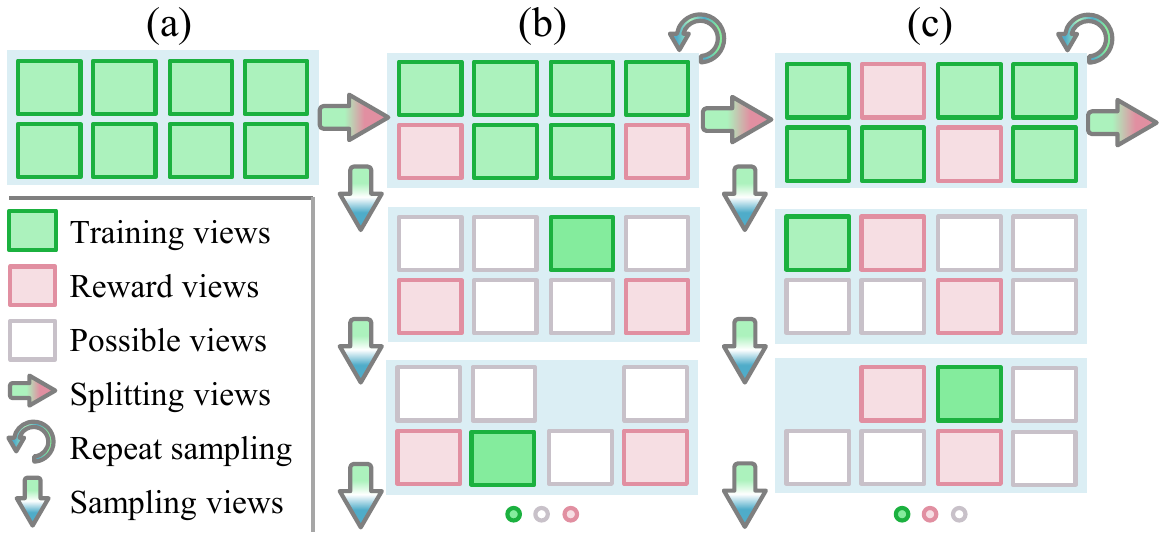}
  \caption{(a) Initial training views.
(b) Split the initial set into training and reward views, then sample training views from the current pool of possible views.
(c) After each $I_{\mathrm{shuffle}}$ operations, resample the reward views.}
  \label{fig:viewsampling}
\end{figure}

To obtain a reward signal that accurately captures multi‑view improvements induced by actions (i.e, changing learning rate or densification hyperparameter), we withhold reward views from the training set. We find that including reward views in the training set introduces a bias toward training views. At the same time, per-scene 3D reconstruction benefits from utilizing as many training views as possible. Balancing between leveraging every view and avoiding over-fitting on bias is non-trivial. 

To address this problem, as shown in Figure~\ref{fig:viewsampling}, we split the initial training views into the reward views and remaining training views. Every $I_{\mathrm{shuffle}}$ iterations, we keep the reward views fixed for short-term reward stability. To leverage all views over the long term, we randomly shuffle and re-split the initial training views into reward views and remaining training views. This design is inspired by cross validation. However, unlike traditional cross-validation, which partitions data for evaluating a fixed hyperparameter configuration, our approach uses the reward views as online feedback to the policy network. The policy thus learns to generate hyperparameters on the fly, directly influencing the 3DGS training process rather than merely evaluating it.

\subsection{3DGS Backbone Design}
We propose a plug-and-play reinforcement learning framework for tuning 3DGS hyperparameters. However, changing hyperparameters inevitably alters the final number of Gaussians, introducing additional challenges for performance assessment. Therefore, we integrate our method into two state-of-the-art 3DGS backbones (TMGS~\cite{mallick2024taming} and 3DGS-MCMC~\cite{kheradmand20243d}) that can set final number of Gaussians.

\begin{table*}[ht]
  \centering
  \small
  \begin{tabular}{l c| c c c | c c c | c c c}     
    \toprule

    \multirow{4}{*}{\textbf{Method}}
      & \multirow{4}{*}{\textbf{num of trials}$\downarrow$}
      & \multicolumn{3}{c|}{\textbf{TNT}}
      & \multicolumn{3}{c|}{\textbf{Deep Blending}}
      & \multicolumn{3}{c}{\textbf{Average}} \\
    \cmidrule(lr){3-5} \cmidrule(lr){6-8} \cmidrule(lr){9-11}
      & 
      & \textbf{PSNR}$\uparrow$
      & \textbf{SSIM}$\uparrow$
      & \textbf{LPIPS}$\downarrow$
      & \textbf{PSNR}$\uparrow$
      & \textbf{SSIM}$\uparrow$
      & \textbf{LPIPS}$\downarrow$
      & \textbf{PSNR}$\uparrow$
      & \textbf{SSIM}$\uparrow$
      & \textbf{LPIPS}$\downarrow$ \\
    \midrule
    TMGS         & 1   & 24.04 & 0.851 & 0.170 & 30.14 & 0.907 & 0.235 & 27.09 & 0.879 & 0.203 \\ 
    \midrule
    RS + TMGS    & 64  & 24.15 & 0.860 & 0.162 & 29.43 & 0.905 & 0.244 & 26.79 & 0.883 & 0.203 \\
    BO + TMGS    & 64  & 24.48 & 0.861 & 0.162 & 29.59 & 0.906 & 0.242 & 27.04 & 0.884 & 0.202 \\
    \midrule
    Ours + TMGS  & \textbf{1}   & \textbf{24.74} & \textbf{0.866} & \textbf{0.158} & \textbf{30.26} & \textbf{0.911} & \textbf{0.233} & \textbf{27.50} & \textbf{0.889} & \textbf{0.195} \\

    \bottomrule
  \end{tabular}
  \caption{Comparison of different methods on the TNT~\cite{knapitsch2017tanks} and Deep Blending datasets~\cite{hedman2018deep}. The last three columns show the average of PSNR, SSIM, and LPIPS across both datasets.}
  \label{tab:exp-search}
\end{table*}

\section{Experiments}
We begin with an overview of datasets and evaluation metrics in Section~\ref{sec:datasets-metrics}. Next, we compare our approach with hyperparameter search methods on public benchmark datasets in Section~\ref{sec:exp-param-search}. Then, we validate our method's effectiveness on a large-scale real-world dataset in Section~\ref{sec:exp-dl3dv}. In Section~\ref{sec:exp-mcmc}, we apply our method to another Gaussian splatting backbone to demonstrate its adaptation. Section~\ref{exp:ablation} presents an ablation study of our method’s key components. Finally, implementation details can be found in Section~\ref{exp:settings}.

\begin{figure*}[!ht]
  \centering
  \includegraphics[width=1.0\linewidth]{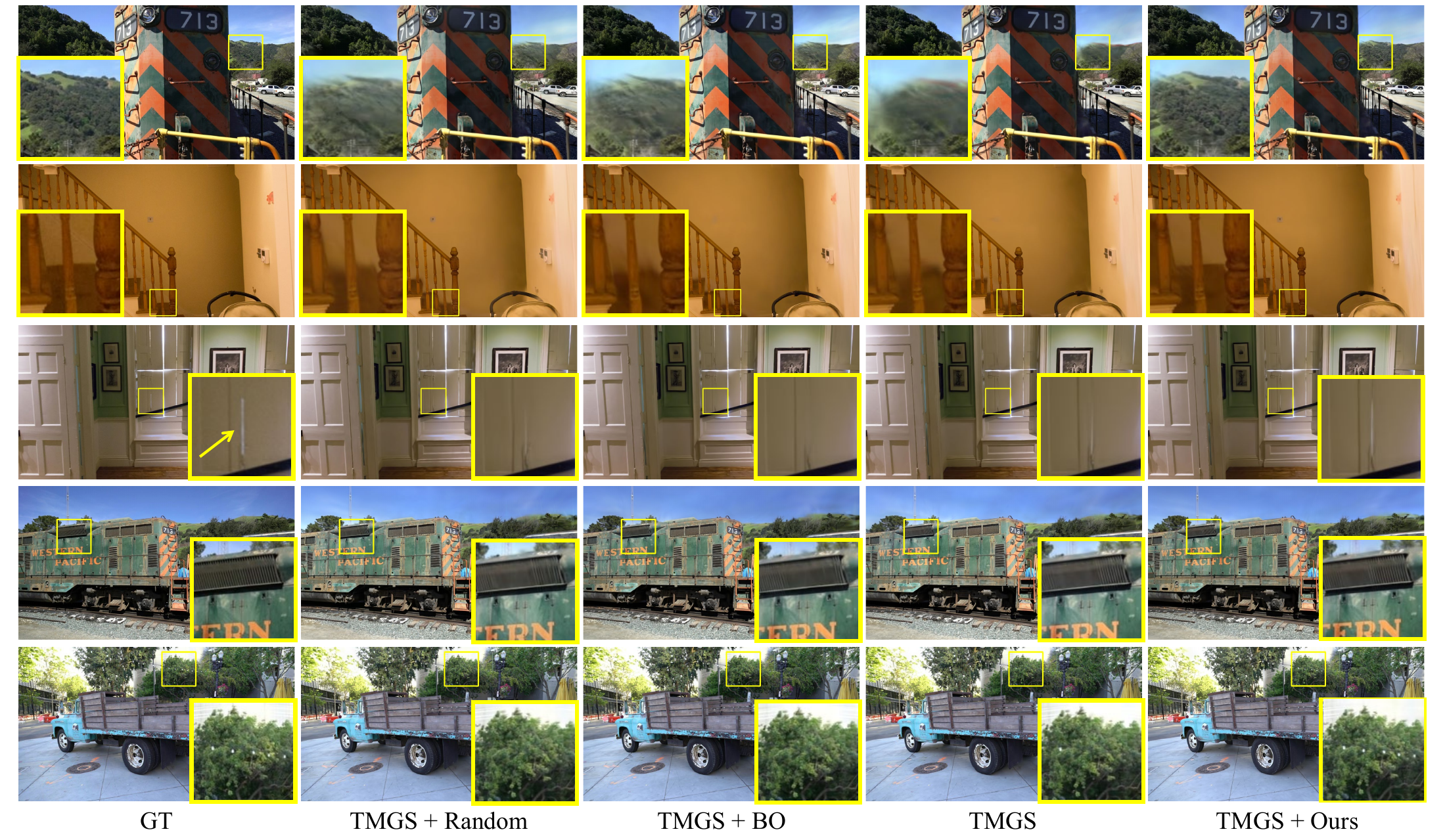}
  \caption{Qualitative comparisons on the TNT and Deep Blending Datasets.}
  \label{fig:comparesearch}
\end{figure*}

\subsection{Datasets and Metric Settings}
\label{sec:datasets-metrics}

\paragraph{TNT and Deep Blending.}
Due to the high computational cost of baselines such as random search, we conduct experiments on two real-world datasets: the Tanks and Temples dataset (TNT)~\cite{knapitsch2017tanks} and the Deep Blending dataset~\cite{hedman2018deep}. Each dataset contains two scenes. These datasets include bounded indoor and unbounded outdoor environments with rich background details.

\paragraph{DL3DV-140.}
To support large-scale tuning across diverse real-world scenes, we evaluate on the DL3DV-140 dataset~\cite{ling2024dl3dv}, which consists of 140 scenes with varied content, created for the novel view synthesis task.

\paragraph{Mip-NeRF360.}
To evaluate the generality of our method, we additionally test on the Mip-NeRF360 dataset~\cite{barron2022mip}. Following the protocol of~\cite{kheradmand20243d}, we select seven representative scenes and downsample the test images for evaluation.

\paragraph{Train/test split.}
For all the datasets, we follow prior work~\cite{kerbl3Dgaussians, mallick2024taming} and reserve every 8th image for testing.

\paragraph{Metrics.}
We report PSNR, SSIM, and LPIPS~\cite{zhang2018perceptual} following standard practice in 3DGS papers~\cite{kerbl3Dgaussians, mallick2024taming}.

\subsection{Comparison with Hyperparameter Tuning Methods}

\label{sec:exp-param-search}
In this experiment, we evaluate our policy–gradient tuner against two standard hyper-parameter optimization (HPO) techniques. We use the Taming-3DGS as the backbone.

\paragraph{Baselines. }  
\emph{Random Search} (\textbf{RS})~\cite{bergstra2012random} samples hyperparameters uniformly at random.
\emph{Bayesian Optimisation} (\textbf{BO})~\cite{bergstra2011algorithms}: we employ the Tree-structured Parzen Estimator (TPE) sampler from Optuna~\cite{akiba2019optuna}, which  optimizes the same training-PSNR objective.
All baselines are built on the latest official codebase of TMGS~\cite{mallick2024taming}.All search methods operate over the same hyperparameter ranges. For \emph{Bayesian Optimization}, we provide the official hyperparameter as a good informative prior.
\paragraph{Results. }  
As shown in Table~\ref{tab:exp-search}, our experiments on two benchmark datasets demonstrate that the default TMGS hyperparameters deliver strong performance for the two dataset. Despite conducting 64 trials, baseline methods random search (RS) and Bayesian optimization (BO) fail to outperform these defaults on the Deep Blending dataset. In contrast, our reinforcement learning–based tuner consistently surpasses them. On the TNT dataset, our method can enhance TMGS by a large margin ($0.7 dB $). 
Visual comparisons are shown in Figure~\ref{fig:comparesearch}. In the last example shown in the figure, our method successfully preserves the small light bulb, whereas search-based methods fail to retain it.

\subsection{Generalization to Real-World Large-Scale Dataset}

\begin{table}[!t]
  \centering
  \small
  \begin{tabular}{l | c c c}     
    \toprule
    \textbf{Method}
    & \textbf{PSNR}$\uparrow$
    & \textbf{SSIM}$\uparrow$
    & \textbf{LPIPS}$\downarrow$ \\
    \midrule
    TMGS  & 29.75 &0.909&0.131\\
    \midrule  
    TMGS + Ours  & 29.96 & 0.912 &0.125\\
    \bottomrule
  \end{tabular}
  \caption{Evaluation on the large-scale DL3DV-140 dataset~\cite{ling2024dl3dv}.}
  \label{tab:dl3dv}
\end{table}
\begin{figure}[!t]
  \centering
\includegraphics[width=1.0\linewidth]{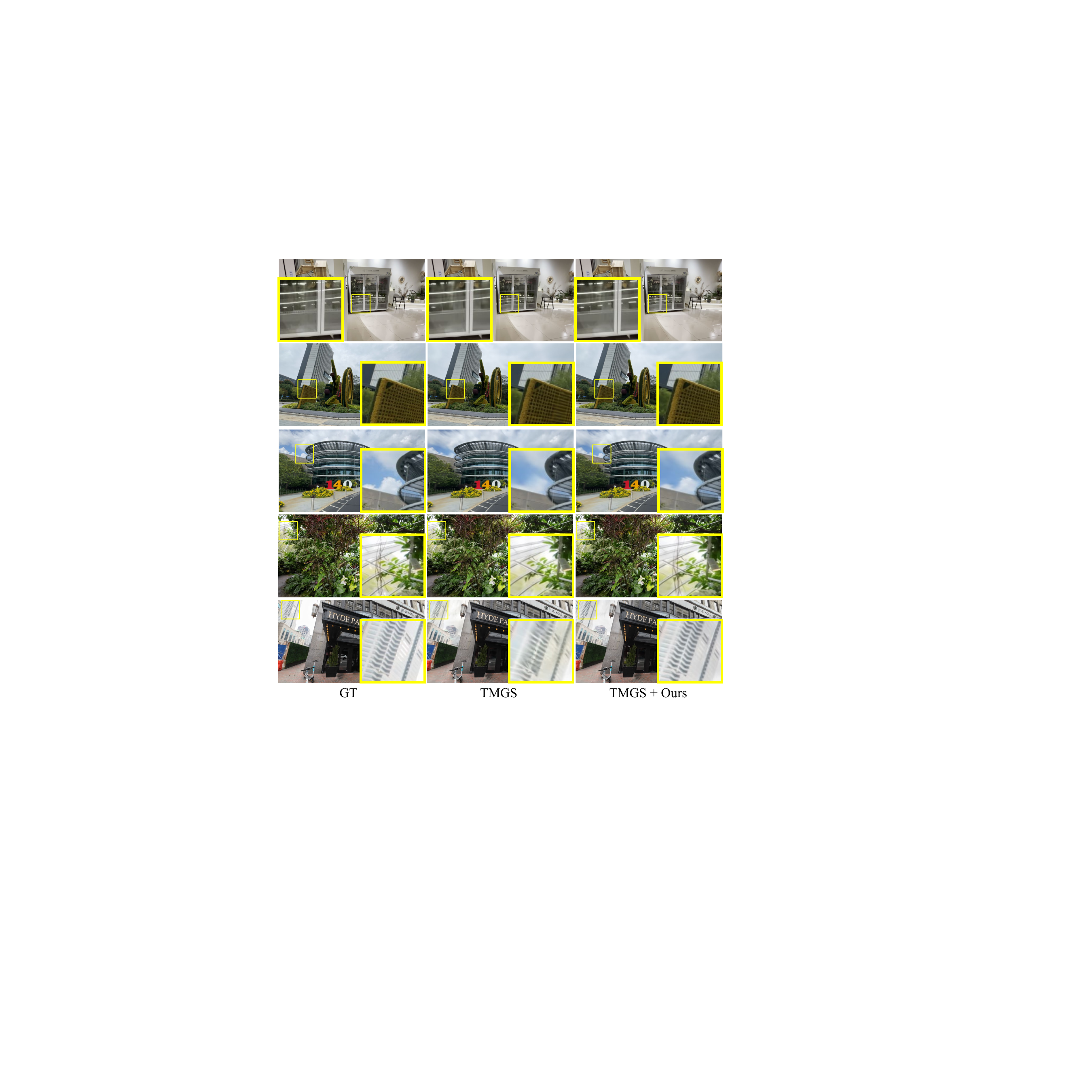}
  \caption{Qualitative comparisons on the DL3DV Datasets.}
  \label{fig:dl3dv}
\end{figure}

\label{sec:exp-dl3dv}
This experiment evaluates our method on DL3DV, a large-scale, user-captured dataset, to assess its adaptability to complex and diverse real-world data domains.

\paragraph{Baseline. } As established in the previous experiment, TMGS is a strong baseline. Therefore, we adopt the state-of-the-art TMGS method as our baseline and integrate our approach into it to evaluate how it can be improved on the DL3DV-140 benchmark.

\paragraph{Results. }
Table~\ref{tab:dl3dv} summarizes quantitative comparisons between TMGS baseline and our adaptive tuning on the DL3DV-140 benchmark. On average across all 140 scenes, our method achieves a 0.21 dB gain in PSNR, and a 0.006 reduction in LPIPS compared to TMGS with its default hyperparameters. These improvements consistently demonstrate the effectiveness of our method over the TMGS backbone. Figure~\ref{fig:dl3dv} provides qualitative examples: while the baseline sometimes blurs thin structures and misses fine textures, our method preserves these details.

\subsection{Adaptation to 3DGS-MCMC Backbone}
\label{sec:exp-mcmc}

\begin{table}[t]
  \centering
  \small
  \begin{tabular}{l | c c c}     
    \toprule
    \textbf{Method}
    & \textbf{PSNR}$\uparrow$
    & \textbf{SSIM}$\uparrow$
    & \textbf{LPIPS}$\downarrow$ \\
    \midrule

    3DGS-MCMC  & 29.89 & 0.900 & 0.190\\
    \midrule

    3DGS-MCMC + Ours  & 30.12 & 0.897 & 0.144 \\
       
    \bottomrule
  \end{tabular}
  \caption{Adaption to 3DGS-MCMC on the MipNeRF 360 dataset~\cite{barron2022mip}.}
  \label{tab:mcmc}
\end{table}

\begin{figure}[!t]
  \centering
  \includegraphics[width=1.0\linewidth]{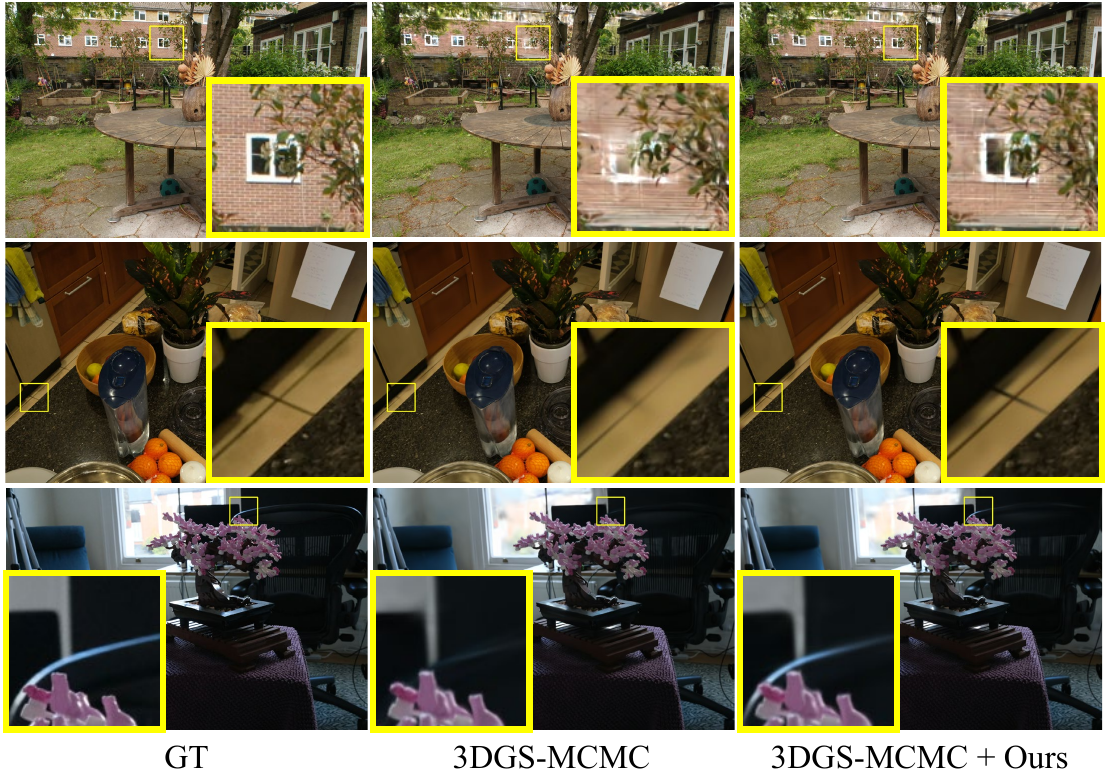}
  \caption{Qualitative comparisons on the MipNeRF 360 Datasets.}
  \label{fig:mcmc}
\end{figure}

\paragraph{Baseline. }  
We also adapt our method to another 3D Gaussian approach to demonstrate its practical value, especially as new 3DGS methods are emerging rapidly. Specifically, we apply our method to 3DGS-MCMC~\cite{kheradmand20243d}, a state-of-the-art technique that also allows control over the number of Gaussians. Since 3DGS-MCMC’s densification process does not involve any tunable hyperparameters—except for the opacity threshold, whose modification would compromise the training process—we replace the densification hyperparameter with the regularization hyperparameter recommended by the authors. This adaptation further demonstrates the flexibility of our framework, which is not limited to tuning the learning rate and densification parameters.

\paragraph{Results. }
Table~\ref{tab:mcmc} presents quantitative comparisons between the 3DGS-MCMC baseline and our adaptively tuned version. Our method yields a 0.23 dB gain in PSNR and a 0.046 (24 \%) reduction in LPIPS. Qualitative results in Figure~\ref{fig:mcmc} further illustrate that our approach better preserves thin structures and fine textures; in the last example, only our method accurately captures the light reflections on the chair. These results demonstrate the generalizability of our framework on other 3DGS variants.

\subsection{Ablation Study}
\label{exp:ablation}
We perform an ablation study on each component of the proposed framework using the TNT dataset (Table~\ref{tab:ablation}). We begin with the two main modules, followed by a detailed analysis of specific design choices, including the GRU encoder, entropy bonus, loss input, and reward-view sampling.

\paragraph{Baseline TMGS. }
"w/o RLLR and RLDS" is the baseline TMGS without our modification. This causes a 0.7 dB PSNR drop, showing that our two modules can significantly enhance reconstruction quality.

\paragraph{Learning rate hyperparameters. }
“w/o RLLR” replaces our reinforcement learning–based learning rate policy with the original fixed schedule. This variant suffers a 0.41 dB drop in PSNR, along with corresponding degradations in SSIM and LPIPS, indicating that our $\mathrm{RLLR}$ module plays a significant role in the overall performance improvement.

\paragraph{Densification hyperparameters. }
“w/o RLDS” replaces our densification hyperparameters policy with the original one in TMGS. Omitting RLDS yields a 0.09 dB PSNR reduction, indicating the effectiveness of our $\mathrm{RLDS}$ module.

\paragraph{Recurrent module architecture. }
“w/o GRU” replaces the GRU encoder with linear layers. This change incurs a 0.11 dB PSNR loss, confirming that GRU better captures long-term information than linear layers.

\paragraph{Entropy term in reward computing. }
“w/o Entropy” removes the entropy term from the policy reward, discouraging exploration. This leads to a 0.12 dB PSNR decrease, suggesting that the entropy bonus helps the policy avoid suboptimal behaviors.

\paragraph{Loss input. }
“w/o Loss Input” excludes the previous-phase reconstruction loss from the policy’s state input. This causes a 0.09 dB PSNR drop, highlighting that feeding back loss information stabilizes policy updates and improves decision making.

\paragraph{Reward view sampling. }
“w/o Reward sampling” uses a reward set from each phase's training set instead of sampling from a hold reward set. This incurs a 0.11 dB PSNR loss, demonstrating that diversity in reward sampling aids generalization across viewpoints.

Overall, both the policy modules ($\mathrm{RLLR}$ and $\mathrm{RLDS}$) and the design choices (GRU encoder, entropy bonus, loss input, and reward-view sampling) contribute significantly to the improved rendering quality.

\begin{table}[ht]
  \centering
  \small
  \begin{tabular}{l | c c c}     
    \toprule
    \textbf{Method}
    & \textbf{PSNR}$\uparrow$
    & \textbf{SSIM}$\uparrow$
    & \textbf{LPIPS}$\downarrow$ \\
    \midrule
    Ours          & 24.74 & 0.866 & 0.158 \\
    \midrule 
    w/o RLLR and RLDS & 24.04 &  0.851  & 0.170 \\
    \midrule
    w/o RLLR     & 24.33 &  0.861& 0.161 \\
    w/o RLDS    & 24.65 & 0.863 & 0.158 \\
    \midrule
    w/o GRU           & 24.63& 0.864 & 0.159 \\
    w/o Entropy        & 24.62 & 0.863 & 0.162 \\
    w/o Loss input      & 24.65 & 0.864 & 0.160 \\
    w/o Reward sampling & 24.63 &0.863 & 0.161 \\
    \bottomrule
  \end{tabular}
  \caption{Ablation study on the TNT dataset.}
  \label{tab:ablation}
\end{table}

\subsection{Implementation Details}
\label{exp:settings}
To stabilize policy training, we clip gradients to a maximum norm of 2.4. The reward-set length is set to 2, and each policy phase spans $K = 20$ training steps. We empirically set \textbf{$I_{\mathrm{shuffle}}$} to 1000: smaller values destabilize the reward signal, while larger values reduce view diversity over time. The learning rate for our policy networks is fixed at $1 \times 10^{-4}$. The \textit{Train} scene takes ~25 minutes to complete training using our integrated TMGS model on an NVIDIA A6000 Ada GPU. The learning-rate policy module $\mathrm{RLLR}$ controls five hyperparameters related to 3DGS optimization: position, scaling, rotation, opacity, and the base spherical harmonic feature coefficients. The densification policy module $\mathrm{RLDS}$ adjusts two key hyperparameters: the density threshold and the external scaling factor used during densification. We do not modify the opacity threshold in the densification process, as changing it would significantly alter the number of Gaussians and potentially break the program. Please refer to the supplemental material for additional implementation details and extended results.

\section{Conclusions}
In this work, we propose a reinforcement learning–based framework (RLGS) for adjusting hyperparameters during 3DGS training. By formulating hyperparameter tuning as an online decision-making problem, our plug-and-play method uses lightweight policies that observe photometric error and training progress to dynamically adjust learning rates and densification hyperparameters.

RLGS integrates seamlessly into existing 3DGS pipelines without requiring architectural changes. Extensive experiments demonstrate its effectiveness: on the TNT dataset, RLGS improves PSNR by up to 0.7 dB. On the large-scale DL3DV-140 dataset, it delivers consistent improvements across diverse real-world scenes and achieves better qualitative results. Furthermore, we show that RLGS generalizes well to 3DGS variants, including Taming-3DGS and 3DGS-MCMC.

A current limitation is that the learned policy network functions as a black box, making it difficult to interpret decisions. Future work will explore more interpretable architectures, incorporate hybrid reinforcement learning techniques, and broaden the scope of adaptive control to other aspects of 3DGS optimization.

\bibliographystyle{2026}

\bibliography{main.bbl}

\appendix
\pagestyle{plain}     
\appendix

\addcontentsline{toc}{section}{Appendix}
\clearpage

\section{Overview}

In the supplementary material, we provide:
\begin{itemize}
     \item The 3DGS preliminary in \textbf{Appendix~\ref{app:preliminary}}.
    \item More qualitative comparisons in \textbf{Appendix~\ref{app:exp-qualitive}}.
    \item Per-scene quantitative results in \textbf{Appendix~\ref{app:exp-perscene}}.
\end{itemize}

\section{3DGS Preliminary}
\label{app:preliminary}
Given multi-view captured images with poses, 3D Gaussian Splatting~\cite{kerbl3Dgaussians} (3DGS) learns a set of anisotropic 3D Gaussians by minimizing a rendering loss through differentiable rasterization. 
In 3DGS~\cite{kerbl3Dgaussians}, each Gaussian $i$ is defined by its center $\mu_i$, covariance $\Sigma_i$, opacity $\sigma_i$, and spherical harmonics coefficients $\mathbf{h}_i$. 
The opacity contributed by Gaussian $i$ at an arbitrary point $\mathbf{x}$ is given by:
\begin{equation}
\label{eq:alpha}
\alpha_i = \sigma_i \exp\left( -\frac{1}{2} (\mathbf{x} - {\mu_i})^T \Sigma_{i}^{-1} (\mathbf{x} - \mu_i) \right).
\end{equation}
Each gaussian's covariance $\Sigma_{i}$ is positive semi-definite and we factor it into a rotation matrix $R_i$ and a diagonal scale matrix $S_i$:
\begin{equation}
\label{eq:covariance}
    \Sigma_i = R_iS_iS_i^TR_i^T.
\end{equation}

To rasterize a 2D image from a given viewpoint, 3D Gaussian splatting approximate the perspective projection of each 3D Gaussian into the image plane~\cite{zwicker2001ewa}. Specifcially, a 3D Gaussian with parameters \((\mu_i,\Sigma_i)\) is approximated by a 2D Gaussian with mean \(\mu_i^{2D}\) and covariance \(\Sigma_i^{2D}\). Given the world‑to‑camera extrinsic \(W\) and the intrinsic matrix \(K\), these are computed as
\begin{align}
\label{eq:proj_mu}
    \mu_i^{2D}&=(K((W\mu_i)/(W\mu_i)_z))_{1:2}, \\
    \label{eq:proj_cov}
    \Sigma_i^{2D}&=(JW\Sigma_i W^TJ^T)_{1:2,1:2},
\end{align}
where $J$ is the Jacobian of the projective transformation.
The subscript \((\cdot)_{1:2}\) selects the x and y components of the projected mean, while \((\cdot)_{1:2,1:2}\) extracts the 2D spatial covariance in the image plane.
After sorting the Gaussians in increasing depth order for ordered volumetric rendering, a pixel's color is computed as:
\begin{equation}
  \mathbf{I} 
  = \sum_{i \in \mathcal{N}} 
    \mathbf{c}_i \,\alpha_i^{2D}
    \prod_{j < i} \bigl(1 - \alpha_j^{2D}\bigr)\,,
  \label{eq:fsplat}
\end{equation}
where $\alpha_i^{2D}$ denotes the 2D variant of Eq.~\eqref{eq:alpha}, modified by replacing 
$\mu_i,\Sigma_i,\mathbf{x}$ with 
$\mu_i^{2D},\Sigma_i^{2D},\mathbf{x}^{2D}$ (the pixel coordinate). 
The term $\mathbf{c}_i$ denotes the RGB color produced by evaluating the spherical harmonics with the view direction (from Gaussian means $\mu_i$ to camera centers) and coefficients.

\section{Additional Qualitative Comparisons}
\begin{figure*}[!ht]
  \centering
  \includegraphics[width=1.0\linewidth]{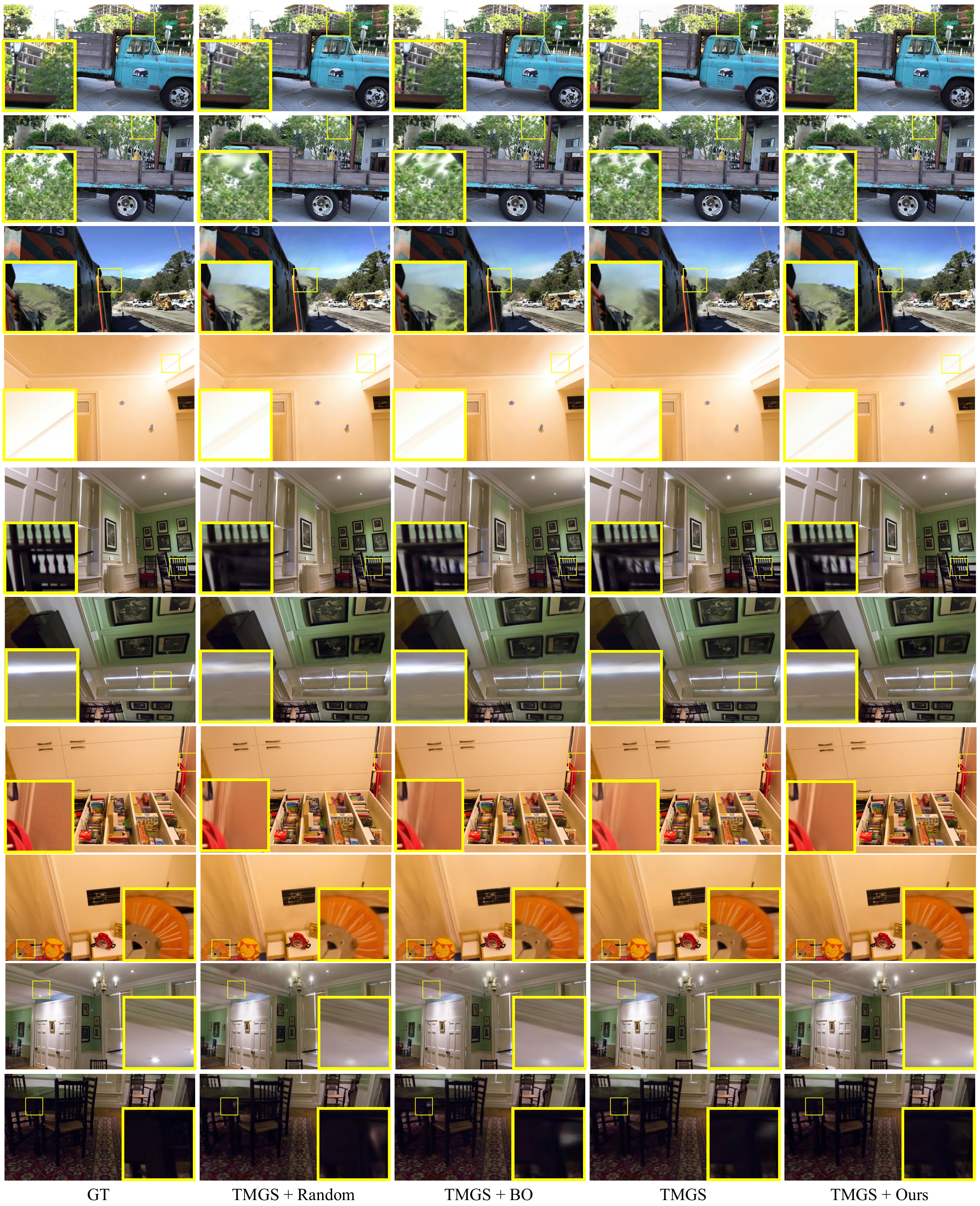}
  \caption{Qualitative comparisons on the TNT and Deep Blending Datasets.}
  \label{fig:app-figure_tntdb}
\end{figure*}

\begin{figure*}[!ht]
  \centering
  \includegraphics[width=0.56\linewidth]{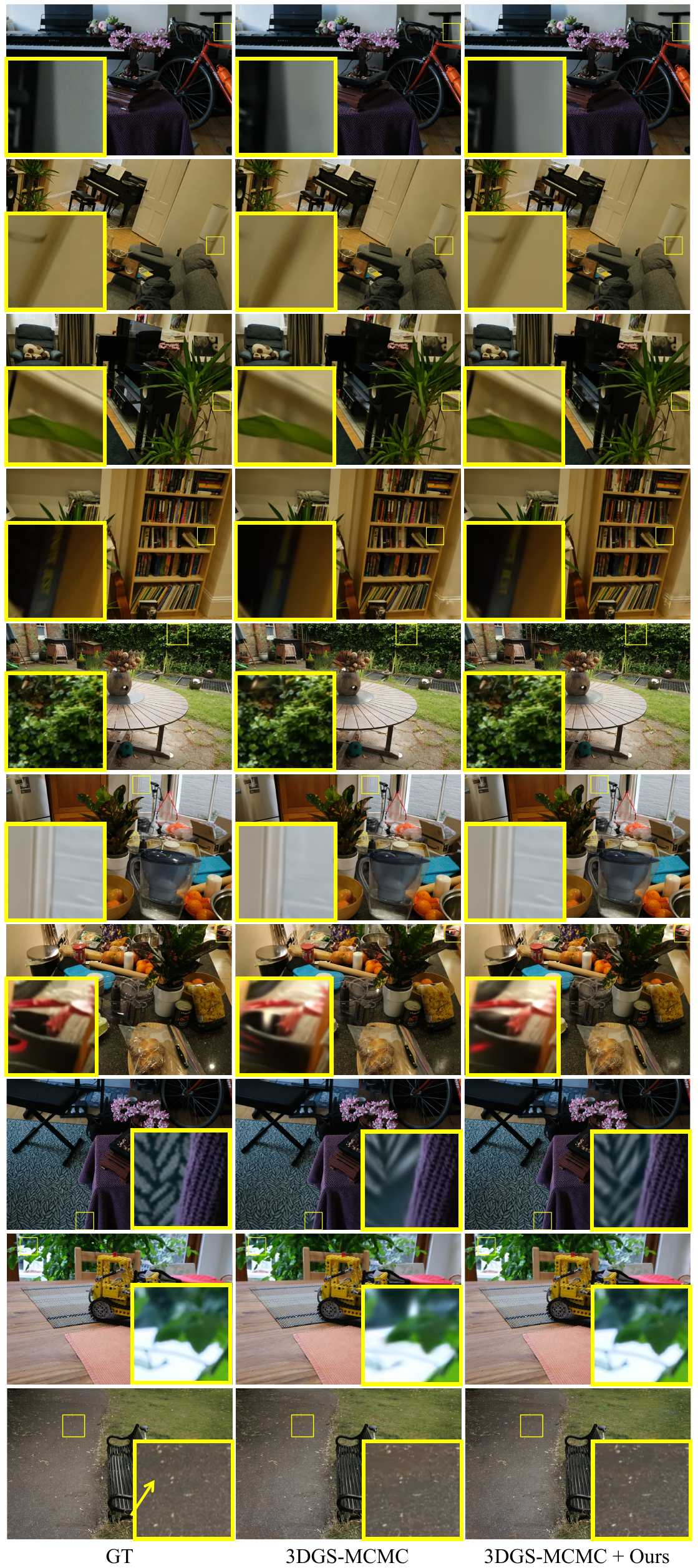}
  \caption{Qualitative comparisons on the MipNeRF 360 Dataset.}
  \label{fig:app-figure_mcmc}
\end{figure*}

\begin{figure*}[!ht]
  \centering
  \includegraphics[width=0.55\linewidth]{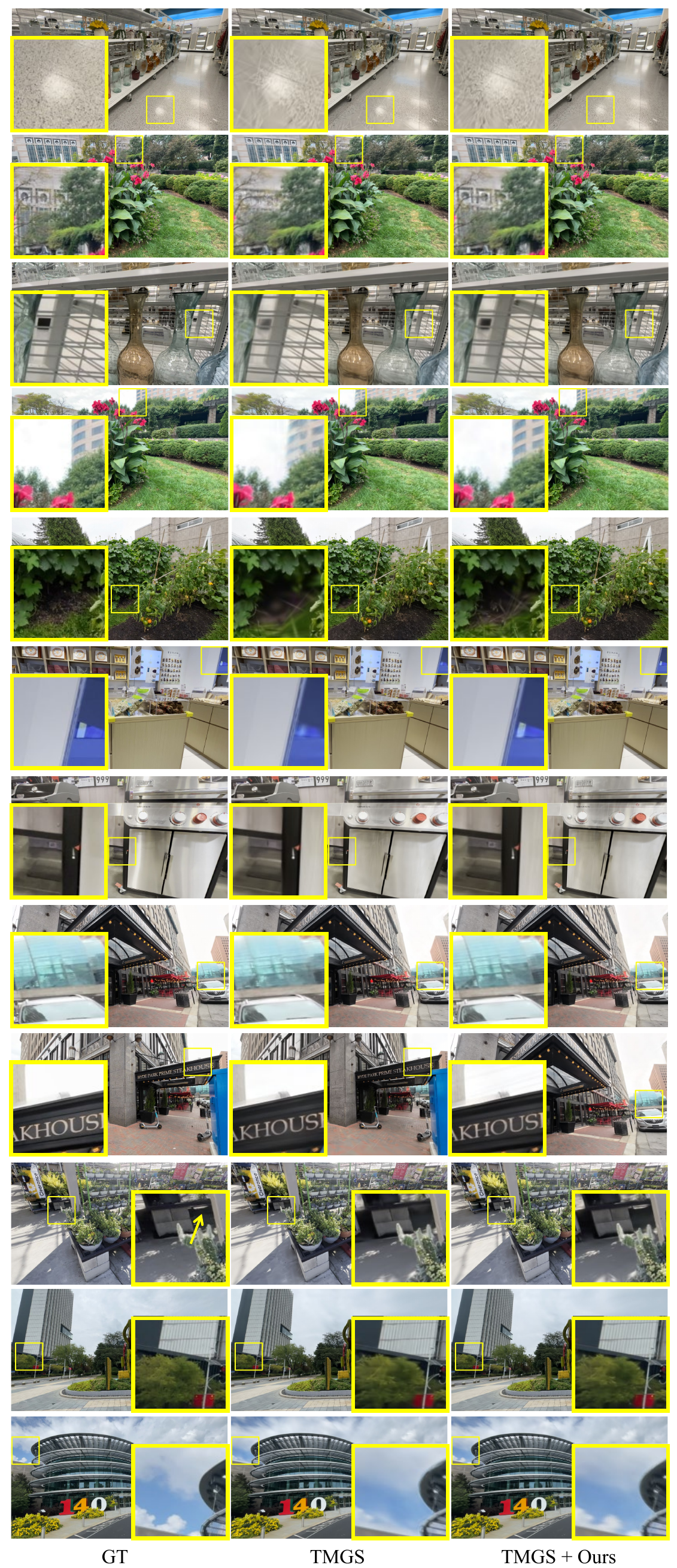}
  \caption{Qualitative comparisons on the DL3DV Dataset.}
  \label{fig:app-figure_dl3dv}
\end{figure*}

In this section, we present additional qualitative comparisons to demonstrate the visual improvements achieved by our method. Figure~\ref{fig:app-figure_tntdb} shows results on the TNT and Deep Blending Datasets. We observe that our method produces noticeably better visual quality in many scenes from the Deep Blending dataset. For example, in the fourth row, our method correctly preserves the edge detail in the brightly lit ceiling, which is not captured as well by the baselines. Figure~\ref{fig:app-figure_mcmc} shows results on the MipNeRF 360 Dataset. When integrated with our method, 3DGS-MCMC preserves better details. In the second-to-last row, which contains challenging blurriness, our method produces results that are nearly indistinguishable from the ground truth. Figure~\ref{fig:app-figure_dl3dv} shows results on the DL3DV Dataset, demonstrating consistent improvements in line with the previous two figures. In the first row of Figure~\ref{fig:app-figure_dl3dv}, where many small spots are affected by ground reflections, our method can reconstruct the rich textures.

\label{app:exp-qualitive}

\section{Per-scene Quantitative Results}
\label{app:exp-perscene}
Table~\ref{tab:app-tnt} shows per-scene quantitative results produced by TMGS~\cite{mallick2024taming} integrated with our method on the TNT~\cite{knapitsch2017tanks} Dataset. Table~\ref{tab:app-db} shows per-scene quantitative results produced by TMGS~\cite{mallick2024taming} integrated with our method on the Deep Blending Dataset~\cite{hedman2018deep}. Table~\ref{tab:app-mip} shows per-scene quantitative results produced by 3DGS integrated with our method. For per-scene results on the DL3DV dataset~\cite{ling2024dl3dv}, please refer to the attached \textit{dl3dv.txt} file.

\begin{table}[ht]
\centering
\begin{tabular}{lcccc}
\toprule
\textbf{Scene}        & \textbf{PSNR} & \textbf{SSIM} & \textbf{LPIPS} & \textbf{Count (M)} \\
\midrule
train & 23.24 & 0.837 & 0.191 & 1.09 \\
truck & 26.25 & 0.895 & 0.124 & 2.58 \\
\bottomrule
\end{tabular}
\caption{Per-scene Quantitative results on the TNT Dataset~\cite{knapitsch2017tanks} using TMGS~\cite{mallick2024taming} enhanced with our method.}
\label{tab:app-tnt}
\end{table}

\begin{table}[ht]
\centering
\begin{tabular}{lcccc}
\toprule
\textbf{Scene} & \textbf{PSNR} & \textbf{SSIM} & \textbf{LPIPS} & \textbf{Count (M)} \\
\midrule
drjohnson & 29.88 & 0.910 & 0.232 & 3.27 \\
playroom  & 30.65 & 0.913 & 0.234 & 2.33 \\
\bottomrule
\end{tabular}
\caption{Per-scene quantitative results on the Deep Blending Dataset~\cite{hedman2018deep} using TMGS~\cite{mallick2024taming} enhanced with our method.}
\label{tab:app-db}
\end{table}

\begin{table}[ht]
\centering
\begin{tabular}{lcccc}
\toprule
\textbf{Scene} & \textbf{PSNR} & \textbf{SSIM} & \textbf{LPIPS} & \textbf{Count (M)} \\
\midrule
bicycle & 26.30 & 0.810 & 0.163 & 5.90 \\
garden  & 28.36 & 0.887 & 0.087 & 5.20 \\
stump   & 27.60 & 0.818 & 0.166 & 4.75 \\
room    & 32.99 & 0.940 & 0.168 & 1.50 \\
counter & 29.70 & 0.927 & 0.159 & 1.20 \\
kitchen & 32.64 & 0.941 & 0.105 & 1.80 \\
bonsai  & 33.26 & 0.955 & 0.160 & 1.30 \\
\bottomrule
\end{tabular}
\caption{Per-scene quantitative results on the MipNeRF 360~\cite{barron2022mip} dataset using 3DGS-MCMC~\cite{kheradmand20243d} enhanced with our method.}
\label{tab:app-mip}
\end{table}

\end{document}